\documentclass{desyproc}

\begin{document}

\title{Photon interactions and chiral dynamics}

\author{{\slshape Wojciech Broniowski$^{1,2}$, Alexandr E. Dorokhov$^3$, Enrique Ruiz Arriola$^4$}\\[1ex]
$^1$Institute of Physics, Jan Kochanowski University, PL-25406~Kielce, Poland\\
$^2$The H. Niewodnicza\'nski Institute of Nuclear Physics, Polish Academy of Sciences, \\PL-31342 Krak\'ow, Poland\\
$^3$Joint Institute for Nuclear Research, Bogoliubov Laboratory of Theoretical Physics, \\114980, Moscow region, Dubna, Russia\\
$^4$Departamento de F\'{\i}sica At\'omica, Molecular y Nuclear, Universidad de Granada, \\E-18071 Granada, Spain }

\contribID{50}

\confID{1407}  
\desyproc{DESY-PROC-2009-03}
\acronym{PHOTON09} 
\doi  

\maketitle

\begin{abstract}
Twist-2 components of the real and virtual photon distribution amplitudes are 
evaluated in several chiral quark models. 
The results, obtained at the quark model scale, are then evolved to higher scales, probed in experiments 
or in lattice QCD. We also analyze the related form factors and coupling constants. Our results are a genuine 
dynamical prediction, following from the chiral dynamics.
\end{abstract}

\section{Basics}

This talk is based on Ref.~\cite{Dorokhov:2006qm}, where more details and results can be found. 
Our approach is based on the fact that the spontaneously broken chiral symmetry provides the basic dynamics 
for the evaluation of soft matrix element involving the Goldstone bosons (pion, kaons) and gauge currents 
(photons, $W^\pm$, $Z$). That way one may evaluate in a genuinely dynamical way the soft quantities 
appearing in high-energy processes. A detailed presentation of the method and the compilation of predictions
for the pion matrix elements can be found in Ref.~\cite{Broniowski:2007si}. 

A crucial ingredient of the method 
is the QCD evolution from the a priori unknown {\em quark model scale} to the scales relevant for the experiments 
of lattice calculations. Thus the scheme consists of two steps: 1)~the evaluation of soft matrix elements in the chiral quark model and 
2)~the QCD evolution to a higher scale. The quark model scale may be estimated with the help of 
the momentum sum rule \cite{Broniowski:2007si}, and is found to be low, $Q_0\simeq 320$~MeV (for the local chiral quark models). 
After the QCD evolution, a successful description of the available data for the pion is achieved for the parton distribution
function (PDF) and the distribution amplitude (DA). There are numerous quark-model 
studies of these quantities 
as well as the more general pion generalized parton distributions (GPD's) in the literature 
\cite{Davidson:1994uv,Dorokhov:1998up,Polyakov:1998td,Polyakov:1999gs,Bakulev:2000eb,%
Dorokhov:2000gu,Anikin:2000sb,Anikin:2000th,RuizArriola:2001rr,%
Davidson:2001cc,Praszalowicz:2001wy,RuizArriola:2002bp,RuizArriola:2002wr,Tiburzi:2002kr,Tiburzi:2002tq,Theussl:2002xp,%
Praszalowicz:2003pr,Broniowski:2003rp,Bzdak:2003qe}. 
A related quantity, the pion-photon transition distribution amplitude (TDA) \cite{Pire:2004ie,Pire:2005ax} has also been evaluated in this 
framework~\cite{Broniowski:2007fs,Courtoy:2007vy,Kotko:2008gy}.

The hadronic part of the photon wave-function consist, in the large-$N_c$ limit, of the quark-antiquark pair. Since the chiral dynamics 
provides the quarks a large (constituent) mass, it influences
the photon dynamics. Here we apply the methods developed and tested earlier for the pion 
to the photon case. We focus on the photon DA's, while the photon structure function is left for a separate study.
The leading-twist photon distribution amplitudes (DA's) are defined via the 
matrix elements of quark bilinears delocalized along the light cone \cite{Ali:1995uy,Ball:1996tb,Ball:1998sk,Ball:2002ps},
\begin{eqnarray}
&&  \langle0|\overline{q}(z)\sigma_{\mu\nu}[z,-z]q(-z)|\gamma^{\lambda}(q)\rangle= \nonumber \\ &&
\hspace{5mm} ie_{q}\langle\bar{q}q\rangle \textcolor{red}{\chi_{\mathrm{m}}}\textcolor{blue}{f_{\perp\gamma}^{t}\left( q^{2}\right)}  
\left(  \epsilon_{\perp\mu}^{(\lambda)}p_{\nu}-\epsilon_{\perp\nu}^{(\lambda)}p_{\mu}\right)  
\int_{0}^{1}dxe^{i(2x-1) q\cdot z}\textcolor{red}{\phi_{\perp\gamma}(x,q^{2})}+ h.t., \nonumber \\
&&  \langle0|\overline{q}(z)\gamma_{\mu}[z,-z]q(-z)|\gamma^{\lambda}(q)\rangle
= \nonumber \\ && \hspace{5mm} e_{q}\textcolor{red}{f_{3\gamma}}\textcolor{blue}{f_{\parallel\gamma}^{v}\left(  q^{2}\right)}  
 p_{\mu }\left(  \epsilon^{(\lambda)}\cdot n\right)  \int_{0}^{1}dxe^{i(2x-1) q\cdot z}\textcolor{red}{\phi_{\parallel\gamma}(x,q^{2})}+ h.t., 
\nonumber
\end{eqnarray}
where $\epsilon^{(\lambda)}\cdot q=0$ and $\epsilon^{(\lambda)}\cdot n=0$ (for real photons) and
\begin{eqnarray}
p_{\mu}=q_{\mu}-\frac{q^{2}}{2}n_{\mu},\;n_{\mu}=\frac{z_{\mu}}{p\cdot z}, \; e_{\mu}^{(\lambda)}=\left(  e^{(\lambda)}\cdot n\right)  p_{\mu}+\left(e^{(\lambda)}\cdot p\right)  n_{\mu}+e_{\perp\mu}^{(\lambda)}. \nonumber
\end{eqnarray}
The quark magnetic susceptibility, $\chi_{\mathrm{m}}$,  and $f_{3\gamma}$ are constants, $f_{\perp\gamma}^{t}(q^{2})$ and 
$f_{\parallel\gamma}^{v}(q^2)$ are form factors, $\phi_{\perp\gamma}(x,q^{2})$ and $\phi_{\parallel\gamma}(x,q^{2})$ denote the DA's, while 
{\em h.t.} stands for the disregarded higher-twist contributions.

\begin{wrapfigure}{r}{0.45\textwidth}
\centerline{\includegraphics[width=0.45\textwidth]{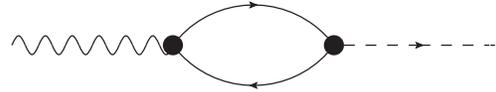}}
\caption{Feynman diagram for the evaluation of the photon DA's in chiral quark models. \label{fig:diag}} 
\end{wrapfigure}

The leading-$N_c$ quark model evaluation proceeds according to the
one-loop diagram, where one of the vertices corresponds to the photon
and the other to the probing operator, in our case $\sigma^{\mu \nu}$
and $\gamma^\mu$.  The quark propagators involve a constituent quark
mass, due to spontaneous breaking of the chiral symmetry.  We use a
few variants of chiral quark models: the Nambu--Jona-Lasinio
(NJL)~(for reviews see, {\em e.g.},
\cite{Christov:1995vm,RuizArriola:2002wr} and references therein) and
the Spectral Quark model
(SQM)~\cite{RuizArriola:2001rr,RuizArriola:2003bs}, which incorporates
the vector-meson dominance, as well as the instanton-motivated
non-local chiral quark model of
Ref.~\cite{Terning:1991yt,Holdom:1992fn,Plant:1997jr,Broniowski:1999bz,Anikin:2000rq}.
In nonlocal models the quark mass depends on the virtuality, As a
consequence, the vertices acquire corrections due to nonlocalities to
consistently account for gauge and chiral Ward identities.

\section{Results}

\begin{table}[tb]
\caption{The constants obtained in the quark model and evaluated to the reference scale of 1~GeV. \label{tab:const}}
\begin{center}
\begin{tabular}{|c|c|c||c|c|}
\hline
quantity at 1~GeV & non-local & SQM & QCD s.r. & VMD \\ \hline 
$ (-\left\langle 0\left\vert \overline{q}q\right\vert 0\right\rangle)^{1/3}$ [GeV]  & 0.24 & 0.24  & 0.24 $\pm$ 0.02  & - \\ 
$\chi_{m}$ [GeV$^2$]                                                                & 2.73 & 1.37  & 3.15 $\pm$ 0.3  & 3.37 \\
$f_{3\gamma}$ [GeV$^{-2}$]                                                          & -0.0035 & -0.0018 & -0.0039 $\pm$ 0.0020 & -0.0046 \\ 
\hline
\end{tabular}
\end{center}
\end{table}

The result for the constants are presented in Table~\ref{tab:const}, where we also give the estimates of the QCD sum rules  and
the Vector Meson Dominance model \cite{Ball:2002ps}. QCD predicts the {\em scale dependence} for the quark condensate 
$\left\langle 0\left\vert \overline{q}q\right\vert 0\right\rangle$, its magnetic
susceptibility $\chi_{\mathrm{m}}$, and $f_{3\gamma}$. At the leading order
\begin{eqnarray}
&& \left.  \left\langle 0\left\vert \overline{q}q\right\vert 0\right\rangle
\right\vert _{\mu}=L^{-\gamma_{\overline{q}q}/b}\left.  \left\langle
0\left\vert \overline{q}q\right\vert 0\right\rangle \right\vert _{\mu_{0}%
}, \; \left.  \chi_{m}\right\vert _{\mu}=L^{-\left(  \gamma_{0}%
-\gamma_{\overline{q}q}\right)  /b}\left.  \chi_{m}\right\vert _{\mu_{0}%
},\quad\left.  f_{3\gamma}\right\vert _{\mu}=L^{-\gamma_{f}/b}\left.
f_{3\gamma}\right\vert _{\mu_{0}} \nonumber
\end{eqnarray}
where $r=\alpha_{s}\left(  \mu^{2}\right)  /\alpha_{s}\left(  \mu_{0}%
^{2}\right)  $, $b=\left(  11N_{c}-2n_{f}\right)  /3$, is the evolution ratio, with
$\gamma_{\overline{q}q}=-3C_{F}$, $\gamma_{0}=C_{F}$, $\gamma_{f}=3C_{A}-C_{F}/3$,  $C_{F}=4/3$, and $C_{A}=3$ for $N_{c}=3$.
We evolve from the quark model scale, $Q_0=320$~MeV, to the reference scale of 1~GeV.
We note a similar magnitude and signs compared to the QCD sum rules or VMD estimates, with the local model producing smaller 
values from the nonlocal model.

\begin{figure}[tb]
\centerline{\includegraphics[width=0.43\textwidth]{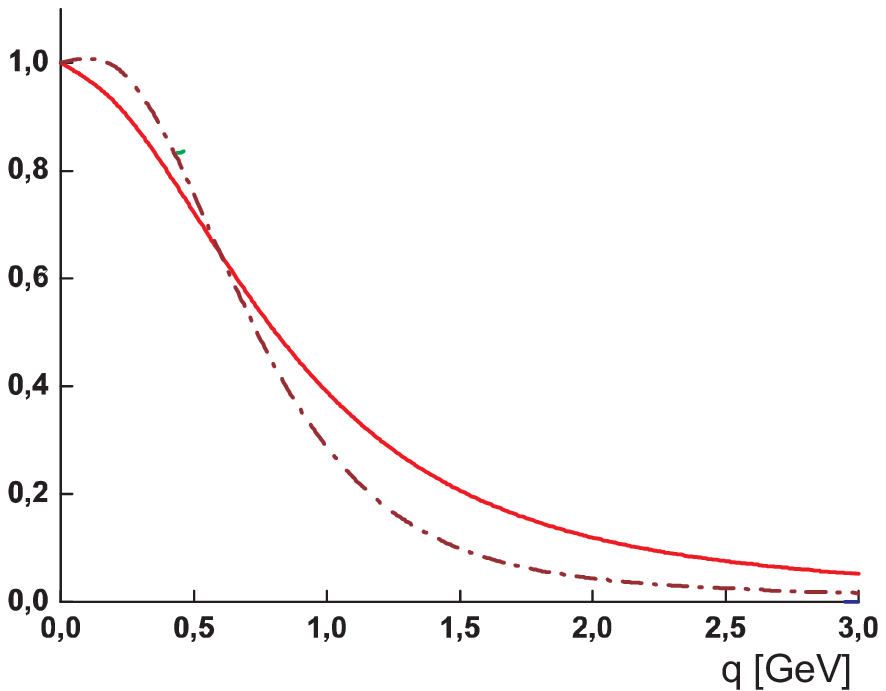} 
\hspace{7mm} \includegraphics[width=0.43\textwidth]{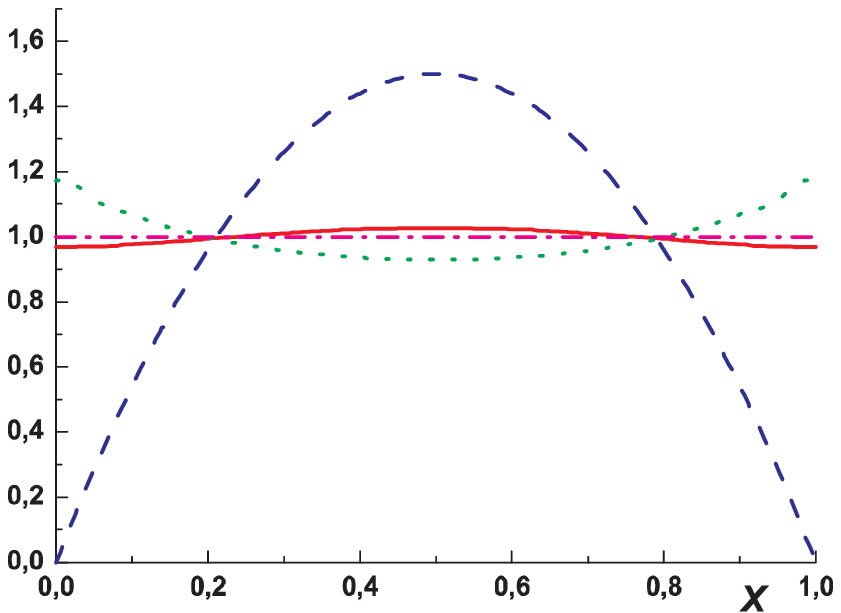}}
\caption{Left: the twist-2 tensor (dashed line) and vector (solid line) form factors in the non-local model. \label{fig:ffphiT}
Right: the transverse DA of the photon, $\phi_{\perp \gamma}(x,q^2=0)$. Solid -- non-local model, dot-dashed -- local model, 
dotted -- approximation of Ref.~\cite{Petrov:1998kg}, dashed -- the asymptotic form $6x(1-x)$.} 
\end{figure}

The form factors from the non-local quark model are shown in the left panel of  
Fig.~\ref{fig:ffphiT}. For the local models (not displayed) the results are very similar. 
They exhibit the typical fall-off scale of $\sim m_\rho$
In particular, in SQM we recover the exact VMD formula 
\begin{eqnarray}
f^{t,{\rm SQM}}_{\perp\gamma }(q^2)=\frac{m_\rho^2}{m_\rho^2+q^2}. \nonumber
\end{eqnarray}

We note that the vector DA $\phi_{\perp \gamma}(x,q^2=0)=1$ in local
models and is very close to $1$ in non-local models.  For the virtual
photon SQM gives the simple formulas:
\begin{eqnarray} 
\phi_{\parallel \gamma^\ast}(x,q^2)= \frac{1+\frac{q^2}{m_\rho^2}}{\left ( 1\!+\! \frac{4q^2}{m_\rho^2}x(1-x)\right )^{3/2}}. \nonumber
\end{eqnarray}
In the limit of $q^2 \to -m_\rho^2$ it becomes $\delta \left (
x-\frac{1}{2} \right )$, a quite reasonable result.

One may also study the photon light-cone wave function (a
$k_\perp$-unintegrated object). It has a simple form in SQM {(at the
  quark-model scale)}:
\begin{equation}
\Phi_{\perp\gamma}%
(x,\mathchoice{\mbox{\boldmath$\displaystyle k$}} {\mbox{\boldmath$\textstyle k$}} {\mbox{\boldmath$\scriptstyle k$}} {\mbox{\boldmath$\scriptscriptstyle k$}}_{\perp
})=\frac{6}{m_{\rho}^{2}%
(1+4\mathchoice{\mbox{\boldmath$\displaystyle k$}} {\mbox{\boldmath$\textstyle
k$}} {\mbox{\boldmath$\scriptstyle k$}} {\mbox{\boldmath$\scriptscriptstyle
k$}}_{\perp}^{2}/m_{\rho}^{2})^{5/2}} \nonumber
\end{equation}
Note the power-law fall-off at large transverse momenta, $\Phi_{\perp\gamma
}%
(x,\mathchoice{\mbox{\boldmath$\displaystyle k$}}{\mbox{\boldmath$\textstyle
k$}} {\mbox{\boldmath$\scriptstyle k$}}{\mbox{\boldmath$\scriptscriptstyle
k$}}_{\perp})\sim1/k_{\perp}^{5}$. In cross section this leads to tails $\sim1/k_{\perp}^{10}$.
For the virtual photon 
\begin{equation}
\Phi_{\perp\gamma^{\ast}}(x,\mathchoice{\mbox{\boldmath$\displaystyle
k$}} {\mbox{\boldmath$\textstyle k$}} {\mbox{\boldmath$\scriptstyle k$}} {\mbox{\boldmath$\scriptscriptstyle k$}}_{\perp
})=\frac{6\left(  1+\frac{q^{2}}{m_{\rho}^{2}}\right)  }{m_{\rho}^{2}\left(
1+4\frac
{\mathchoice{\mbox{\boldmath$\displaystyle k$}} {\mbox{\boldmath$\textstyle
k$}} {\mbox{\boldmath$\scriptstyle k$}} {\mbox{\boldmath$\scriptscriptstyle
k$}}_{\perp}^{2}+q^{2}x(1-x)}{m_{\rho}^{2}}\right)  ^{5/2}} \nonumber
\end{equation}

\section{QCD evolution of DA's}

\begin{figure}[tb]
\begin{center}
\includegraphics[width=0.47\textwidth]{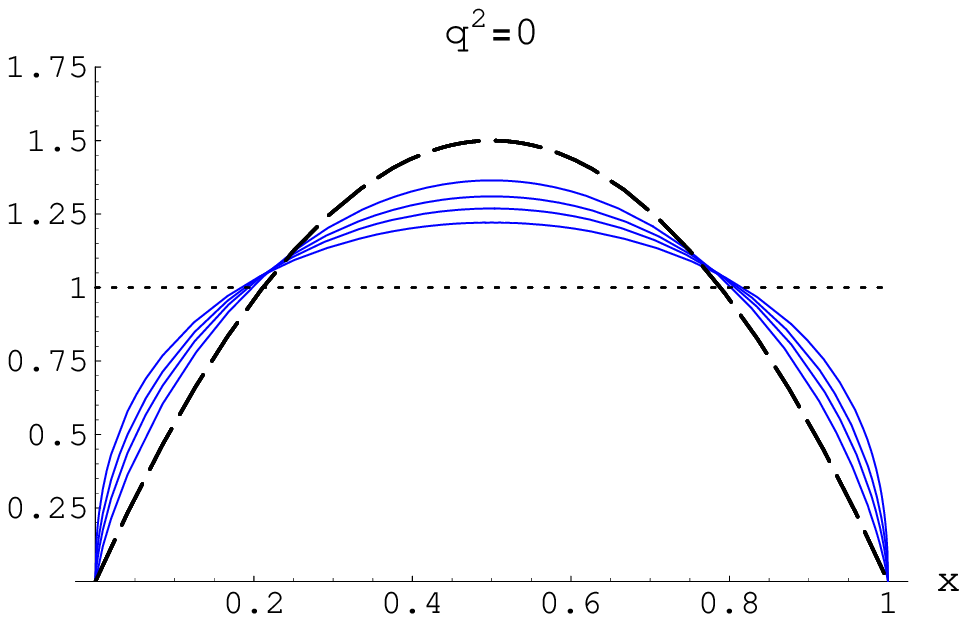} \hspace{3mm} 
\includegraphics[width=0.47\textwidth]{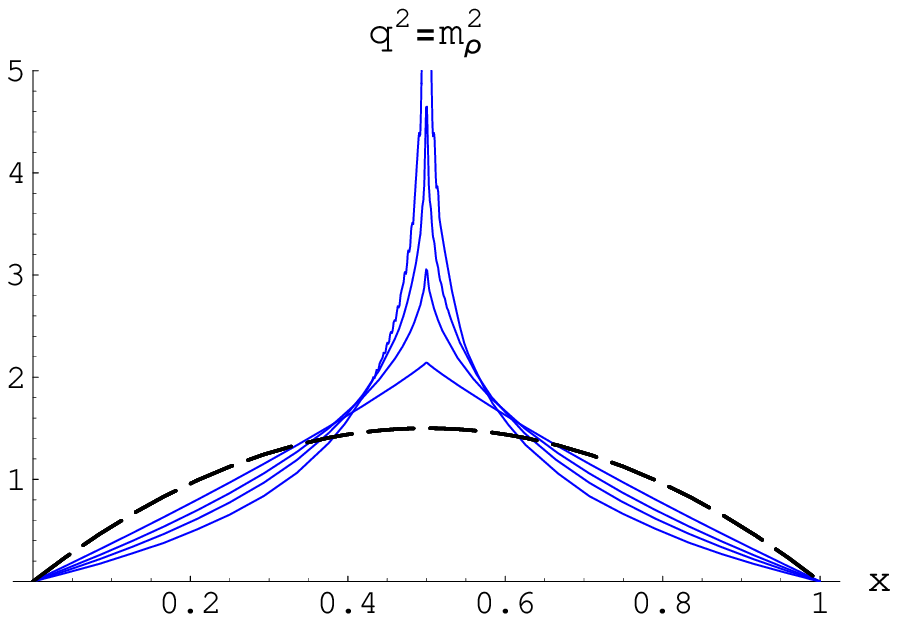} \\
\includegraphics[width=0.47\textwidth]{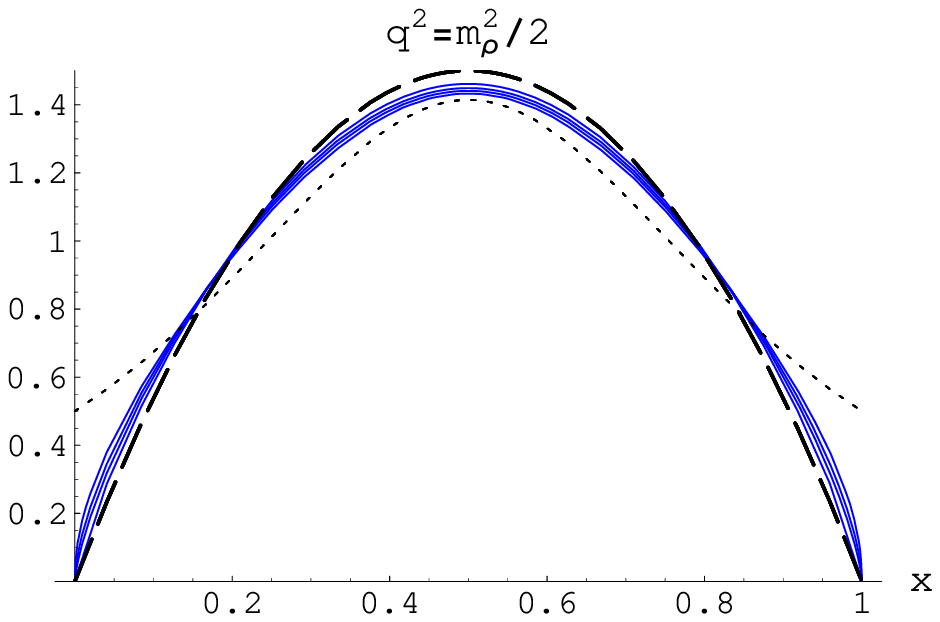} \hspace{3mm} 
\includegraphics[width=0.47\textwidth]{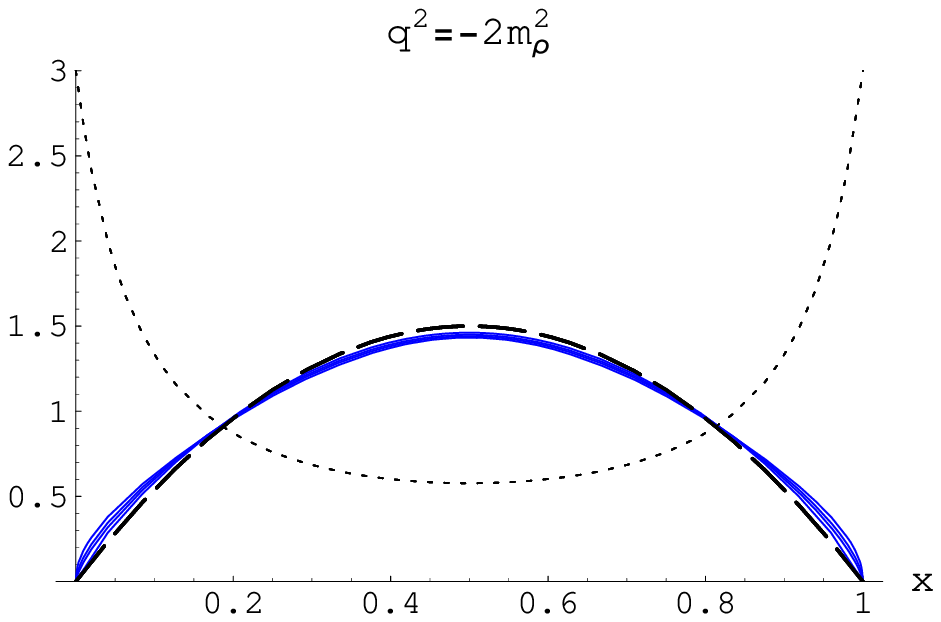}
\end{center}
\caption{The leading order ERBL evolution of the leading-twist tensor DA, $\phi_{\perp\gamma}(x,q^{2})$ 
evaluated in the local model at various virtualities: real photon (top left), $\rho$-meson (top
right), virtual photon at $q^{2}=-Q^2=m_{\rho}^{2}/2$ (bottom left), and virtual
photon at $q^{2}=-Q^2=-2m_{\rho}^{2}$ (bottom right). Initial conditions, indicated
by dotted lines, are evaluated in SQM at the initial
quark-model scale. The solid lines correspond to LO QCD evolution
to the scales $Q=1$, $2.4$, $10$, and $1000$~GeV. With the larger the scale
the evolved DA becomes closer to the asymptotic form $6x(1-x)$, plotted with the
dashed line. The corresponding values of the evolution ratio $r$
are given in the figures. \label{fig:evol}} 
\end{figure}

\begin{figure}[tb]
\begin{center}
\includegraphics[angle=0,width=0.47\textwidth]{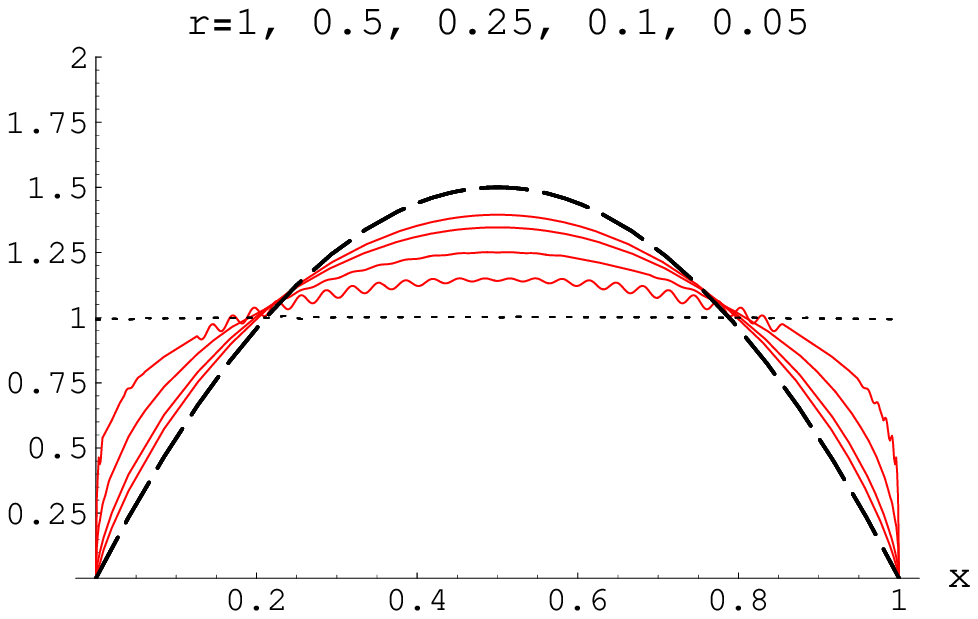} 
\includegraphics[angle=0,width=0.47\textwidth]{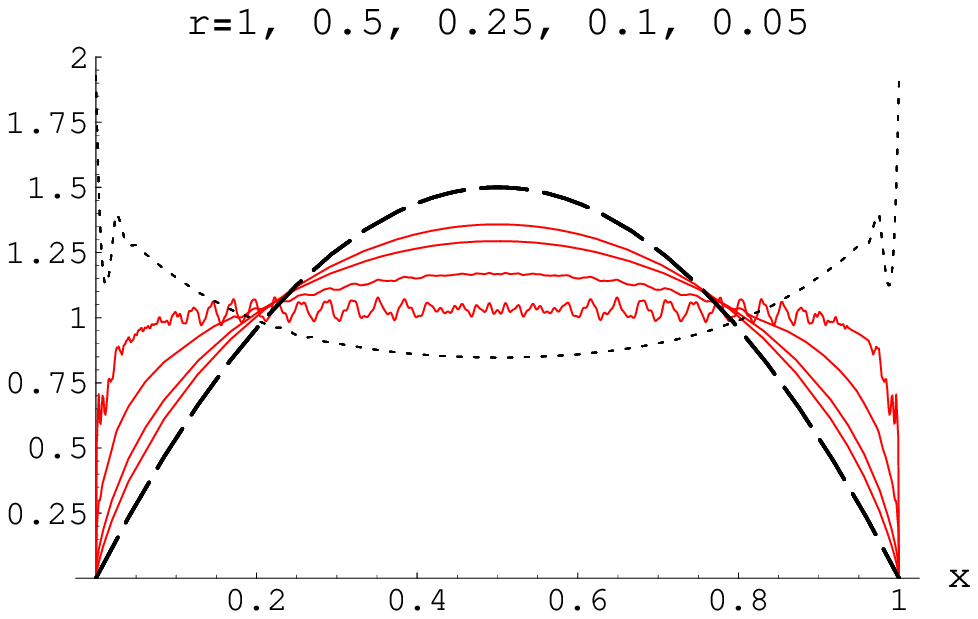}
\end{center}
\caption{The LO ERBL evolution of the nonlocal model predictions for the
leading-twist {vector} DAs of the photon, $\phi_{\parallel\gamma}(x,q^{2})$. Left: real photon,  right: the virtual photon at
$q^{2}=0.25$ GeV$^{2}$. The dashed lines show the asymptotic DA, $6x(1-x)$.
The initial conditions (dotted line) are evaluated in the nonlocal
quark model at the initial scale $Q_0^{\mathrm{inst}}=530$~MeV. The solid
lines correspond to evolved DA'a at subsequent scales $Q=1$, $2.4$, $10$, and $1000$~GeV. 
The corresponding values of the evolution ratio $r$
are given in the figures. Tiny wiggles in the evolved curves is a numerical effect.
}%
\label{NLQMevV}%
\end{figure}

Now we come to the QCD evolution, which, as already stressed, is crucial in bringing the results to the 
scales probed in experiments or lattices. We carry out the standard LO ERBL
evolution with anomalous dimensions taken for the appropriate channels \cite{Belitsky:2005qn}. The method  
leads to simple expressions, diagonal in the
Gegenbauer moments.  In Fig.~\ref{fig:evol} we show the results for
the tensor DA for the real photon, the $\rho$-meson, and the virtual photon. We note the large
change caused by the evolution, which fairly fast brings the model predictions to the vicinity of the 
asymptotic limit. Similar results can be done in the
nonlocal model, as well as for the vector DA \cite{Dorokhov:2006qm}. We show the results in Fig.~\ref{NLQMevV}.

\section{Conclusion}

Chiral quark models provide a link between high- and low-energy
analyses, allowing to compute various soft matrix elements for
hadronic processes.  They yield in a fully dynamical way the initial
conditions for the QCD evolution, which is {essential} to bring the
predictions up to the experimental or lattice scales.  Numerous
predictions for processes involving the Goldstone bosons and photons
can be made.  The scale in chiral quark models is low, about 320~MeV,
hence the QCD evolution is ``fast''.  Simple analytic formulas --
useful to understand the general properties, can be obtained in local
quark models.  For the pion, with the LO QCD evolution the overall
agreement with the available data and lattice simulations is {very
  reasonable} (PDF, DA, GPD, TDA, generalized form factors
\cite{Broniowski:2008hx}).  While the presented results for the
can be used in phenomenological analyses in high-energy
reactions (see, {\em e.g.}, the recent work of Ref.~\cite{Pire:2009ap}), the model predictions can be further tested also with future
lattice simulations for the photon and $\rho$-meson.

\section*{Acknowledgments}

This work has been supported in part by the JINR Bogolyubov-Infeld program, 
the Scientific School grant 4476.2006.2
the Polish Ministry of Science and Higher Education grants
N202~034~32/0918 and N202~249235, and by the Spanish DGI and FEDER
funds with grant FIS2008-01143/FIS, Junta de Andaluc{\'\i}a grant
FQM225-05, and EU Integrated Infrastructure Initiative Hadron Physics
Project contract RII3-CT-2004-506078.

\begin{footnotesize}

\end{footnotesize}

\end{document}